\documentstyle[aps,epsfig]{revtex}
\begin{document}
\preprint{} \draft

\title{The Universe as a Nonuniform Lattice in the Finite-Dimensional
Hypercube.
\\II.Simple Cases of Symmetry Breakdown and Restoration} \thispagestyle{empty}

\author{A.E.Shalyt-Margolin\hspace{1.5mm}\thanks
{Fax: (+375) 172 326075; e-mail: a.shalyt@mail.ru;alexm@hep.by}}
\date{}
\maketitle
 \vspace{-25pt}
{\footnotesize\noindent  \\
\\National Center of Particles and
High Energy Physics, Bogdanovich Str. 153, Minsk 220040, Belarus\\
{\ttfamily{\footnotesize
}}

\rm\normalsize \vspace{0.5cm}
\begin{abstract}
This paper continues a study of field theories specified for the
nonuniform lattice in the finite-dimensional hypercube with the
use of the earlier described deformation parameters. The paper is
devoted to spontaneous breakdown and restoration of symmetry in
simple quantum-field theories with scalar fields. It is
demonstrated that an appropriate deformation opens up new
possibilities for symmetry breakdown and restoration. To
illustrate, at low energies (far from the Planck's) it offers
high-accuracy reproducibility of the same results as with a
nondeformed theory. In case of transition from low to higher
energies and vice versa it gives description for new types of
symmetry breakdown and restoration depending on the rate of the
deformation parameter variation in time, and indicates the
critical points of the previously described lattice associated
with a symmetry restoration. Besides, such a deformation enables
one to find important constraints on the initial model parameters
having an explicit physical meaning.

\end{abstract}

\section{Introduction}
This paper is written to continue a study presented in
\cite{shal1}. In the earlier work it has been shown that with the
use of a new small parameter $\alpha$,$0<\alpha\leq1/4$ and its
statistical mechanics counterpart $\tau$,$0<\tau\leq1/4$,
resultant from the density matrix deformation in quantum and
statistical mechanics of the Early Universe (Planck scale) and
introduced by the author in his previous works forming a series
\cite{shal1}--\cite{shal6}, the Universe may be described as a
nonuniform lattice in the hypercube with finite edges. It should
be noted that a quantum theory naturally appears in this picture,
the transition to higher energies being nothing else but the
transition to lower lattice nodes. In \cite{shal1},\cite{shal3}one
of the simplest transitions from the density matrix deformation
(density pro-matrix) $\rho(\alpha)$ ($\rho(\tau)$ respectively) to
deformation in a quantum field theory $\psi(\widetilde{\alpha},x)$
, or from Neumann to Schrodinger picture, may be found. The
present paper is a study into a simple case of spontaneous
symmetry breakdown for this deformed quantum theory with scalar
fields. It is demonstrated that at lower energies far from the
Planck's the results of the "nondeformed" case are reproduced with
a very high accuracy. Compared to the nondeformed case, new
results associated with symmetry breakdown and restoration appear
on going from low to higher energies and vice versa due to the
rate of the deformation parameter variation in time. The
transition from high energies to lower ones is of particular
importance, being closely associated with cosmological models
\cite{cosm1},\cite{cosm2}. Points of the symmetry restoration are
marked on the lattice $Lat^{\tau}_{\widetilde{\alpha}}$,introduced
previously in \cite{shal1}, and referred to as the critical
points. The constraints imposed on the deformation parameters lead
to significant additional conditions for the initial model
parameters having an explicit physical meaning.
\\ The paper is structured as follows: section 1 is an introduction
of the problem. Section 2 is a brief outline of the principal
results obtained on the deformed density matrix in quantum
mechanics $\rho(\alpha)$ and statistical mechanics of the Early
Universe $\rho(\tau)$; transition to the associated deformation of
a quantum field theory and deformed wave function
$\psi(\widetilde{\alpha},x)$, giving the basis for the description
of lattice $Lat^{\tau}_{\widetilde{\alpha}}$. The section 3
presents the results obtained on spontaneous symmetry breakdown
and restoration in model with scalar fields. And in Conclusion the
main inferences of the work are summarized.

\section {Problem Survey and Notation}
Now it is commonly accepted that physics of the Early Universe
\cite{Ahl1} (scales on the order of the Planck's) should be
distinct from the physics of the Universe in its current state.
This may be inferred proceeding from different approaches, e.g.,
string theory \cite{Str1}, quantum gravics
\cite{Mag1},\cite{Ahl2},\cite{Gar1} and some others
\cite{Adl1},\cite{Cap}. And quantum mechanics of the Early
Universe may be considered as a deformation of a well-known
quantum mechanics. In this case the deformation is understood as
an extension of a particular physical theory by the introduction
of an additional parameter or several parameters in such a way
that the initial theory appears in the limiting transition to
certain fixed values of these parameters \cite{Fad}. As distinct
from the approach associated with the Heisenberg's algebra
deformation that modifies the Heisenberg uncertainty relations
\cite{Heis}, for example \cite{Mag1}, \cite{Mag2},
\cite{Cap},\cite{KMM}), in his publications
\cite{shal1}--\cite{shal6} the author has approached the
deformation of quantum mechanics of the Early Universe using
radically new method - the density matrix deformation. In his
approach the deformation parameter is a dimensionless value
$\alpha$: $\alpha=l_{min}^{2}/x^{2}$, where $l_{min}$ is a minimal
length and $x$ is the measuring scale. Generally it is accepted
that the value of $l_{min}$ is on the order of the Planck's length
$l_{min} \sim l_{p}$. However, this assumption is not a "must".  A
new deformed object $\rho(\alpha)$, referred to as a {\bf density
pro-matrix} possesses a number of interesting features described
in \cite{shal2},\cite{shal3}, giving in the low-energy limit an
ordinary quantum-mechanical density matrix. Moreover,
$\rho(\alpha)$ allows for a new approach to the solution of some
problems: Liouville equation deformation in the Early Universe and
close to the black hole \cite{shal3}, black hole entropy and its
quantum corrections
\cite{shal1},\cite{shal3}--\cite{shal5},entropy density
\cite{shal3}--\cite{shal5}, information loss problems in the
processes associated with black holes \cite{shal4},\cite{shal5}.
\\ We are especially interested in the following feature of $\rho(\alpha)$:
\begin{equation}\label{U2.1}
Sp[\rho(\alpha)]-Sp^{2}[\rho(\alpha)]\approx\alpha.
\end{equation}
From whence it follows that the value for $Sp[\rho(\alpha)]$ satisfies
the above-stated condition
\begin{equation}\label{U2.2}
Sp[\rho(\alpha)]\approx\frac{1}{2}+\sqrt{\frac{1}{4}-\alpha}
\end{equation}
and therefore  $0<\alpha\leq1/4$.
\\ Besides, solution (\ref{U2.1}) will be further used as an exponential
ansatz \cite{shal2},\cite{shal3}:
\begin{equation}\label{U2.3}
\rho^{*}(\alpha)=\sum_{i}\alpha_{i} exp(-\alpha)|i><i|,
\end{equation}
where all
$\alpha_{i}>0$ are independent of $\alpha$  and their sum is equal
to 1. Then
\begin{equation}\label{U2.4}
Sp[\rho^{*}(\alpha)]=exp(-\alpha).
\end{equation}
In \cite{shal6} it has been demonstrated that statistical
mechanics of the Early Universe should be also "deformed" compared
to a well-known statistical mechanics associated with Gibbs
distribution \cite{Feyn}. Similar to quantum mechanics, the
principal object in this case is just the density matrix, though
now statistical.  Here the deformation parameter is $\tau =
T^{2}/T_{max}^{2}$, varying in the same interval $0<\tau\leq1/4$
as $\alpha$. A new object appearing as a result of this
deformation $\rho_{stat}(\tau)$, that is called the statistical
density pro-matrix, gives in the limit of temperatures far from
maximum $T_{max}\sim T_{p}$ (where $T_{p}$ is Planck's
temperature)a well-known statistical density matrix  $\rho_{stat}$
\cite{Feyn} and satisfies the analog of conditions described by
(\ref{U2.1})-(\ref{U2.4}) with a change of $\alpha$ by $\tau$
\cite{shal6}, i.e.
\begin{equation}\label{U2.1S}
Sp[\rho_{stat}(\tau)]-Sp^{2}[\rho_{stat}(\tau)]\approx \tau.
\end{equation}
Correspondingly,
\begin{equation}\label{U2.2S}
Sp[\rho_{stat}(\tau)]\approx\frac{1}{2}+\sqrt{\frac{1}{4}-\tau}.
\end{equation}
Naturally, with an appropriate change exponential ansatz
(\ref{U2.3}) is here also the case, and solution (\ref{U2.1S})
satisfies the equation
\begin{equation}\label{U2.4S}
Sp[\rho^{*}(\tau)]=exp(-\tau).
\end{equation}
Taking into consideration duality of relation between time and
temperature that follows from the Generalized Uncertainty
Relations in quantum theory and thermodynamics
\cite{shal6},\cite{shal7}:
\begin{equation}\label{U2.5S}
\left\{
\begin{array}{lll}
\Delta x & \geq & \frac{\displaystyle\hbar}{\displaystyle\Delta
p}+\alpha^{\prime} L_{p}^2\,\frac{\displaystyle\Delta
p}{\displaystyle\hbar}+... \\
  &  &  \\
  \Delta t & \geq &  \frac{\displaystyle\hbar}{\displaystyle\Delta E}+\alpha^{\prime}
  t_{p}^2\,\frac{\displaystyle\Delta E}{\displaystyle\hbar}+... \\
  &  &  \\

  \Delta \frac{\displaystyle 1}{\displaystyle T} & \geq &
  \frac{\displaystyle k}{\displaystyle\Delta U}+\alpha^{\prime}
  \frac{\displaystyle 1}{\displaystyle T_{p}^2}\,
  \frac{\displaystyle\Delta U}{\displaystyle k}+...,
\end{array} \right.
\end{equation}
In \cite{shal1} it has been shown that the Universe may be
represented as a four-dimensional lattice
$Lat^{\tau}_{\widetilde{\alpha}}$ in hypercube $I_{1/4}^{4}$ with
edge $I_{1/4}=(0;1/4]$, where an arbitrary point of
$Lat^{\tau}_{\widetilde{\alpha}}$ is characterized by coordinates
$(\widetilde{\alpha},\tau)=
(\alpha_{1},\alpha_{2},\alpha_{3},\tau)$, and
 $\alpha_{1},\alpha_{2},\alpha_{3}$ are the values of parameter $\alpha$
in three space dimensions. They are taken as independent of each
other as, in principle, at very high energies (on the order of the
Planck's) the space coordinates may be noncommutative
\cite{Ahl1},\cite{Ahl2},\cite{Gar1}:
\\
$$\left[ x_i ,x_j \right]\neq 0.$$
\\
In this case, independently of each other, all the described
variables may take on one and the same discrete and nonuniform
series of values $1/4, 1/16, 1/36,1/64,...$.
\\Since for exponential ansatz (\ref{U2.3})a prototype
of the pure state (wave function) in the introduced formalism is
represented by the density pro-matrix
\begin{equation}\label{U2.5}
\rho^{*}(\alpha)=exp(-\alpha)|\psi><\psi|,
\end{equation}
$\alpha$-deformation of wave functions, i.e. fields, in QFT is of
the following form \cite{shal3}:
\begin{equation}\label{U2.6}
\psi(x) \mapsto \psi(\alpha,x)=\mid \theta (\alpha)\mid \psi(x),
\end{equation}
where\begin{equation}\label{U2.7}
\mid \theta (\alpha)\mid = exp(-\alpha/2)
\end{equation}
or
\begin{equation}\label{U2.8}
\theta (\alpha)=\pm exp(-\alpha/2)(cos\gamma \pm isin\gamma).
\end{equation}
Further in the text (section 3) it is assumed that
\\
\\1) $\alpha$ of the exponential factor in formula (\ref{U2.3})
is the same for all space coordinates
$\alpha_{1},\alpha_{2},\alpha_{3}$ of the point of lattice
$Lat^{\tau}_{\widetilde{\alpha}}$. This means that as yet the
noncommutativity effect is of no special importance, and parameter
$\alpha$ is determined by the corresponding energy scale;
\\
\\2) parameter $\alpha$ is dependent on time only
$\alpha=\alpha(t)$. This condition is quite natural since $\alpha$
plays a part of the scale factor and is most often dependent on
time only (especially in cosmology) \cite{cosm1},\cite{cosm2});
\\
\\3) as all physical results should be independent
of a selected normalization $\theta (\alpha)$, subsequent choice
of the normalization will be
\begin{equation}\label{U2.9}
\theta (\alpha)=exp(-\alpha/2),
\end{equation}
that is $\gamma=0$.

\section{Spontaneous Symmetry Breakdown and Restoration
in a Model with Scalar Fields}
 First, take a well-known Lagrangian for a scalar field
\cite{cosm3}:
\begin{equation}\label{U3.1}
L=\frac{1}{2}(\partial_{\mu}\phi)^{2}+\frac{1}{2}\mu^{2}\phi^{2}
-\frac{1}{4}\lambda\phi^{4},
\end{equation}
where $\lambda>0$.
\\ Because of the transformation in the above deformation
\begin{equation}\label{U3.2}
\phi \Rightarrow \phi(\alpha)=exp(-\frac{\alpha}{2})\phi,
\end{equation}
where $\alpha=\alpha(t)$, $\alpha$ is a deformed Lagrangian of the form
\begin{equation}\label{U3.3}
L(\alpha)=\frac{1}{2}(\partial_{\mu}\phi(\alpha))^{2}
+\frac{1}{2}\mu^{2}\phi(\alpha)^{2}
-\frac{1}{4}\lambda\phi(\alpha)^{4}.
\end{equation}
It should be noted that a change from Lagrangian to the
Hamiltonian formalism in this case is completely standard
\cite{Div} with a natural changing of $\phi$ by $\phi(\alpha)$.
Consequently, with the use of a well-known formula
\begin{equation}\label{U3.4}
H(\alpha)=p\dot{q}-L(\alpha)
\end{equation}
we obtain
\begin{equation}\label{U3.5}
H(\alpha)=\frac{1}{2}(\partial_{0}\phi(\alpha))^{2}
+\frac{1}{2}\sum_{i=1}^3 (\partial_{i}\phi(\alpha))^{2}
-\frac{1}{2}\mu^{2}\phi(\alpha)^{2}
+\frac{1}{4}\lambda\phi(\alpha)^{4}.
\end{equation}
We write the right-hand side of (\ref{U3.5}) so as to solve it for
$\phi$ and $\alpha$. As $\alpha$ is depending solely on time, the
only nontrivial term (to within the factor of 1/2) in the
right-hand side of (\ref{U3.5}) is the following:
\\
$$(\partial_{0}\phi(\alpha))^{2}=
exp(-\alpha)(\frac{1}{4}\dot{\alpha}^{2}\phi^{2}
-\dot{\alpha}\phi\partial_{0}\phi+(\partial_{0}\phi)^{2})$$.
\\
Thus, $\alpha$ - deformed Hamiltonian $H(\alpha)$ is rewritten as
\begin{equation} \label{U3.6}
H(\alpha)=exp(-\alpha)[-\frac{1}{2}\dot{\alpha}\phi\partial_{0}\phi
+\frac{1}{2}\sum_{j=0}^3 (\partial_{j}\phi)^{2} \nonumber
\\ +(\frac{1}{8}\dot{\alpha}^{2}-\frac{1}{2}\mu^{2})\phi^{2}
+\frac{1}{4}exp(-\alpha)\lambda\phi^{4}].
\end{equation}
To find a minimum of $H(\alpha)$, we make its partial derivatives
with respect $\phi$ and $\alpha$ equal to zero
\begin{equation}\label{U3.7}
\left\{
\begin{array}{ll}
\frac{\displaystyle\partial H(\alpha)}{\displaystyle\partial
\phi}=exp(-\alpha) [-\frac{1}{2}\dot{\alpha}\partial_{0}\phi
\nonumber
\\ +(\frac{1}{4}\dot{\alpha}^{2}-\mu^{2})\phi
+exp(-\alpha)\lambda\phi^{3}]=0
\\
\frac{\displaystyle\partial H(\alpha)}{\displaystyle\partial
\alpha}=-exp(-\alpha)
[-\frac{1}{2}\dot{\alpha}\phi\partial_{0}\phi
+\frac{1}{2}\sum_{j=0}^3 (\partial_{j}\phi)^{2}+ \nonumber
\\ (\frac{1}{8}\dot{\alpha}^{2}-\frac{1}{2}\mu^{2})\phi^{2}
+\frac{1}{2}exp(-\alpha)\lambda\phi^{4}]=0
\end{array}\right.
\end{equation}
That is equivalent to
\begin{equation}\label{U3.8}
\left\{
\begin{array}{ll}
\frac{\displaystyle\partial H(\alpha)}{\displaystyle\partial
\phi}\sim[-\frac{1}{2}\dot{\alpha}\partial_{0}\phi
+(\frac{1}{4}\dot{\alpha}^{2}-\mu^{2})\phi
+exp(-\alpha)\lambda\phi^{3}]=0
\\
\frac{\displaystyle\partial H(\alpha)}{\displaystyle\partial
\alpha}\sim [-\frac{1}{2}\dot{\alpha}\phi\partial_{0}\phi
+\frac{1}{2}\sum_{j=0}^3 (\partial_{j}\phi)^{2} \nonumber \\
+(\frac{1}{8}\dot{\alpha}^{2}-\frac{1}{2}\mu^{2})\phi^{2}
+\frac{1}{2}exp(-\alpha)\lambda\phi^{4}]=0
\end{array}\right.
\end{equation}
Now consider different solutions for (\ref{U3.7}) or (\ref{U3.8}).
\\
I.In case of deformation parameter $\alpha$ weakly dependent on
time
\begin{equation}\label{U3.9}
\dot{\alpha}\approx 0 .
\end{equation}
Most often this happens at low energies (far from the Planck's).
\\ Actually, as $\alpha=l_{min}^{2}/a(t)^{2}$,
where $a(t)$ is the measuring scale, (\ref{U3.9}) is nothing else
but $\alpha \approx 0$,  and the process takes place at low
energies or at rather high energies but at the same energy scale,
meaning that scale factor  $a(t)^{-2}$ is weakly dependent on
time. The first case is no doubt more real. In both cases,
however, we have a symmetry breakdown and for minimum of
$\tilde{\sigma}$ in case under consideration
\begin{equation}\label{U3.10}
\tilde{\sigma} = \pm\mu\lambda^{-1/2}exp(\alpha/2).
\end{equation}
Based on the results of \cite{shal1}--\cite{shal3}, it follows that
\begin{equation}\label{U3.11}
<0\mid \phi\mid0>_{\alpha}=exp(-\alpha)<0\mid \phi \mid0>,
\end{equation}
and we directly obtain
\begin{equation}\label{U3.11}
<0\mid\tilde{\sigma}\mid0>_{\alpha}=\pm\mu\lambda^{-1/2}exp(-\alpha/2).
\end{equation}
Then in accordance with the exact formula
\cite{cosm3},\cite{Div} we shift the field $\phi(\alpha)$
\begin{equation}\label{U3.12}
\phi(\alpha) \Rightarrow \phi(\alpha)+<0\mid
\tilde{\sigma}\mid0>_{\alpha}=exp(-\alpha/2)(\phi+\sigma)=
\phi(\alpha)+\sigma(\alpha),
\end{equation}
where $\sigma=\pm\mu\lambda^{-1/2}$- minimum in a conventional
nondeformed case \cite{cosm3}. When considering $\alpha$ -
deformed Lagrangian (\ref{U3.3}) for the shifted field
$\phi+\sigma$
\begin{equation}\label{U3.13}
L(\alpha)=L(\alpha,\phi+\sigma),
\end{equation}
we obtain as expected a massive particle the squared mass of that
contains, as compared to a well-known case, the multiplicative
exponential supplement $exp(-\alpha)$:
\begin{equation}\label{U3.14}
m_{\phi}^{2}=2\mu^{2}exp(-\alpha),
\end{equation}
leading to the familiar result for low energies or correcting the
particle's mass in the direction of decreasing values for high
energies.
\\
II. Case of $\dot{\alpha} \neq 0$.
\\ This case is associated with a change from low to higher energies
or vice versa. This case necessitates the following additional
assumption:
\begin{equation}\label{U3.15}
\partial_{0}\phi\approx 0.
\end{equation}
What is the actual meaning of the assumption in (\ref{U3.15})? It
is quite understandable that in this case in $\alpha$ - deformed
field $\phi(\alpha)=exp(-\frac{\alpha(t)}{2})\phi$ the principal
dependence on time $t$ is absorbed by exponential factor
$exp(-\frac{\alpha(t)}{2})$. This is quite natural at sufficiently
rapid changes of the scale (conforming to the energy) that is just
the case in situation under study.
\\ Note that in the process a change from Lagrangian to the
Hamiltonian formalism (\ref{U3.4}) for $\alpha$ - deformed
Lagrangian $L(\alpha)$ (\ref{U3.3}) holds true since, despite the
condition of (\ref{U3.15}), we come to
\begin{equation}\label{U3.16}
\partial_{0}\phi(\alpha)\neq 0
\end{equation}
due to the presence of exponential factor $exp(-\frac{\alpha(t)}{2})$.
\\Proceed to the solution for (\ref{U3.7}) (and respectively (\ref{U3.8})).
Taking (\ref{U3.15}) into consideration, for a minimum of
$\tilde{\sigma}$ in this case we obtain
\begin{equation}\label{U3.17}
\tilde{\sigma} =
\pm(\mu^{2}-\frac{1}{4}\dot{\alpha}^{2})^{1/2}\lambda^{-1/2}exp(\alpha/2).
\end{equation}
So, the requisite for the derivation of this minimum will be as follows:
\begin{equation}\label{U3.18}
4\mu^{2}-\dot{\alpha}^{2}\geq 0
\end{equation}
or with an assumption of $\mu>0$
\begin{equation}\label{U3.19}
-2\mu \leq\dot{\alpha} \leq 2\mu.
\end{equation}
Assuming that $\alpha(t)$ is increasing in time, i.e. on going
from low to higher energies and with $\dot{\alpha}>0$, we have
\begin{equation}\label{U3.20}
 0<\dot{\alpha}\leq 2\mu
\end{equation}
or\begin{equation}\label{U3.21}
\alpha(t) \leq 2\mu t.
\end{equation}
From where it follows that on going from low to higher energies,
i.e. with increasing energy of model(\ref{U3.3}), two different
cases should be considered.
\\
IIa. Symmetry breakdown when $\alpha(t) < 2\mu t$
\\
$$\tilde{\sigma} =
\pm(\mu^{2}-\frac{1}{4}\dot{\alpha}^{2})^{1/2}\lambda^{-1/2}exp(\alpha/2).
$$
\\
IIb. Symmetry restoration when $\alpha(t) = 2\mu t$ as in this case
\\
$$\tilde{\sigma}=0.$$
\\
Because of $\alpha(t) \sim a(t)^{-2}$, these two cases IIa and IIb
may be interpreted as follows: provided $a(t)$ increases more
rapidly than $(2\mu t)^{-1/2}$ (to within a familiar factor), we
have a symmetry breakdown at hand as in the conventional case
(\ref{U3.1}), whereas for similar increase a symmetry restoration
occurs. This means that at sufficiently high energies associated
with scale factor $a(t)\sim (2\mu t)^{-1/2}$ the broken symmetry
is restored.
\\ Provided that in this case the time dependence of $a(t)$
is exactly known, then by setting the equality
\\
$$a(t)=l_{min}(2\mu t)^{-1/2}$$
\\
and solving the above equation we can find $t_{c}$ - critical time
for the symmetry restoration. Then the critical scale (critical
energies)
\\
$$a(t_{c})=l_{min}(2\mu t_{c})^{-1/2},$$
\\
actually the energies whose symmetry is restored, and finally the
corresponding critical point of the deformation parameter
\\
$$\alpha(t_{c})=l_{min}^{2}a(t_{c})^{-2}=2\mu t_{c}.$$
\\
This point is critical in a sense that for all points with the
deformation parameter below its critical value
\\
$$\alpha(t)<\alpha(t_{c})$$
\\
the symmetry breakdown will be observed. Obviously, in case under
study the energy is constantly growing with corresponding lowering
of the scale and hence
\\
$$a(t)\sim t^{\xi},\xi<0.$$
\\
\\ Of particular interest is a change from high to lower energies.
This is associated with the fact that all cosmological models may
be involved, i.e. all the cases where scale $a(t)$ is increased
due to the Big Bang \cite{cosm1},\cite{cosm2}.
\\ For such a change with the assumption that $\alpha(t)$ diminishes
in time and $\dot{\alpha}<0$ we obtain
\begin{equation}\label{U3.22}
-2\mu \leq\dot{\alpha}<0.
\end{equation}
Since by definition $\alpha(t)>0$ is always the case and
considering $\alpha(t)$ as a negative increment (i.e.
$d\alpha(t)<0$), we come to the conclusion that the case under
study is symmetric to IIa and IIb.
\\IIc.  For fairly high energies, i.e. for $\alpha(t) = 2\mu t$
or scale $a(t)$ that equals to $(2\mu t)^{-1/2}$(again to within
the familiar factor), there is no symmetry breakdown in accordance
with case IIb.
\\
\\IId. On going to lower energies associated with $\alpha(t) < 2\mu t$
there is a symmetry breakdown in accordance with case IIa and with
the formula of(\ref{U3.17}). Note that in this case energy is
lowered in time and hence the scale is growing, respectively.
Because of this,
\\
$$a(t)\sim t^{\xi},\xi>0.$$
\\
For the specific cases with exactly known relation between $a(t)$
and $t$ one can determine the points without the symmetry
breakdown. In cosmology \cite{cosm2} in particular we have
\\
\\1) in a Universe dominated by nonrelativistic matter
\\
$$a(t)\propto t^{2/3},$$
\\
i.e.
\\
$$a(t)\approx a_{1}t^{2/3}.$$
\\
From whence at the point of unbroken symmetry
\begin{equation}\label{U3.23}
a_{1}t^{2/3}\approx(2\mu t)^{-1/2},
\end{equation}
directly giving the critical time $t_{c}$
\begin{equation}\label{U3.24}
t_{c}\approx(2\mu)^{-3/7}a_{1}^{-6/7},
\end{equation}
and for $t>t_{c}$ the symmetry breakdown is observed (case IId).
\\ In much the same manner
\\
\\2) in the Universe dominated by radiation
\\
$$a(t)\propto t^{1/2},$$
\\
i.e.
\\
$$a(t)\approx a_{2}t^{1/2}$$
\\
from where for $t_{c}$ we have
\begin{equation}\label{U3.25}
t_{c}\approx (2\mu)^{-1/2}a_{2}^{-1},
\end{equation}
and again for $t>t_{c}$ the symmetry is broken.
\\ Here it is interesting to note that despite the apparent symmetry
of cases IIa,IIb and IId , IIc, there is one important
distinction.
\\ In cases IId and IIc time $t$ is usually (in cosmological
models as well \cite{cosm1},\cite{cosm2}) counted from the Big
Bang moment and therefore fits well to the associated time
(temperature) coordinate of the lattice
$Lat^{\tau}_{\widetilde{\alpha}}$ (section 2 and
\cite{shal1},\cite{shal6}. As a result, when the critical time
$t_{c}$ is known, one can find the critical point of the
above-mentioned lattice as follows:
$(\alpha_{c},\tau_{c})=(\widetilde{\alpha_{c}},\tau_{c})$, where
all the three coordinates of the space part
$Lat^{\tau}_{\widetilde{\alpha}}$, i.e. in
$(\widetilde{\alpha_{c}})$, are equal to
\\
\\$$\alpha_{c}=l_{min}^{2}a(t_{c})^{-2},$$
\\
and
\\
\\$$\tau_{c}=T_{c}^{2}/T_{max}^{2},$$
\\
where $T_{c}\sim 1/t_{c}$ ((\ref{U2.5S}) in section 2 and
\cite{shal6},\cite{shal7}.
\\ Thus, in these cases all points $Lat^{\tau}_{\widetilde{\alpha}}$,
for which the following conditions are satisfied:
\begin{equation}\label{U3.26}
\left\{
\begin{array}{ll}
\alpha<\alpha_{c},
\\
\tau<\tau_{c}
\end{array}\right.
\end{equation}
are associated with a symmetry breakdown, whereas at the critical
point $(\alpha_{c},\tau_{c})$ no symmetry breakdown occurs. As
seen from all the above formulae, the point of the retained
symmetry is associated with higher temperatures and energies than
those (\ref{U3.26}), where a symmetry breakdown takes place, in
qualitative agreement with the principal results of \cite{cosm3}.
\\ Note that for cases IIa and IIb time $t$ is a certain local time
of the quantum process having no direct relation to the time
(temperature) variable of lattice
$Lat^{\tau}_{\widetilde{\alpha}}$. Because of this, it is possible
to consider only the critical value at the space part
$Lat_{\widetilde{\alpha}}$ \cite{shal1} of lattice
$Lat^{\tau}_{\widetilde{\alpha}}$, i.e. $\alpha_{c}\in
Lat_{\widetilde{\alpha}}$, all other inferences remaining true
with a change of (\ref{U3.26}) by
\begin{equation}\label{U3.26A}
\alpha<\alpha_{c}.
\end{equation}
However, provided there is some way to find temperature $T_{c}$
that is associated with a symmetry restoration(e.g, \cite{cosm3}
case IIb), one can directly calculate $\tau_{c}$ and finally
change (\ref{U3.26A})by (\ref{U3.26}).
\\ In any case conditions $0<\alpha\leq 1/4;0<\tau\leq 1/4$
(see section 2 and \cite{shal1}--\cite{shal6}) impose constraints
on the model parameters (\ref{U3.26}) and hence on $\mu$, which
may be found in the explicit form by solving the following
inequality:
\\
\\$$0<\alpha_{c}\leq 1/4.$$

\section{Conclusion}
Thus, from primary analysis of such a simple model as (\ref{U3.1})
it is seen that its $\alpha$ - deformation (\ref{U3.3})
contributes considerably to widening the scope of possibilities
for a symmetry breakdown and restoration.
\\
\\1) At low energies (far from the Planck's) it reproduces with
a high accuracy (up to $exp(-\alpha)\approx 1$)  the results
analogous to those given by a nondeformed theory \cite{cosm3}.
\\
\\2) On going from low to higher energies or vice versa it provides
new cases of a symmetry breakdown or restoration depending on the
variation rate of the deformation parameter in time, being capable
to point to the critical points of the earlier considered lattice
$Lat^{\tau}_{\widetilde{\alpha}}$ , i.e. points of the symmetry
restoration.
\\
\\3) This model makes it possible to find important constraints
on the parameters of the initial model having an explicit physical
meaning.
\\ It should be noted that for $\alpha$ - deformed theory (\ref{U3.3})
 there are two reasons to be finite in ultraviolet:
\\ - cut-off for a maximum momentum $p_{max}$;
\\ - damping factors of the form $exp(-\alpha)$,
where in the momentum representation $\alpha=p^{2}/p_{max}^{2}$ in
each order of a perturbation theory suppressing the greatest
momenta.
\\ As a result, some problems associated with a conventional
theory \cite{cosm3} (divergence and so on) in this case are
nonexistent.
\\ In his further works the author is planning to proceed
to $\alpha$ - deformations (symmetry breakdown and restoration,
critical points an so forth) of more elaborate theories involving
the gauge $A_{\mu}$ and spinor $\psi$ fields.


\end{document}